\documentclass{elsart}

\usepackage{latexsym}
\usepackage{amssymb}
\usepackage{amsfonts}
\usepackage{amsmath}
\usepackage[dvips]{graphicx}
\usepackage{bm}

\def\hb#1{\hat{\mathbf{#1}}}
\def\vb#1{\vec{\mathbf{#1}}}

\def\ber{\begin{eqnarray}}
\def\eer{\end{eqnarray}}
\def\beq{\begin{equation}}
\def\eeq{\end{equation}}

\usepackage{natbib}

\usepackage{epsfig}

\usepackage{amssymb}

\begin{document}

\begin{frontmatter}



\title{A null frame for spacetime positioning by means of pulsating sources}


\author[a,b]{Angelo Tartaglia}
\ead{angelo.tartaglia@polito.it}
\author[a,b]{Matteo Luca Ruggiero \corauthref{cor}}
\corauth[cor]{Corresponding author}
\ead{matteo.ruggiero@polito.it}
\author[a,b]{Emiliano Capolongo}
\address[a]{Dipartimento di Fisica, Politecnico di Torino, Corso Duca degli Abruzzi 24, 10129 Torino, Italy}
\address[b]{INFN, Sezione di Torino, Via Pietro Giuria 1, 10125 Torino, Italy}
\ead{emiliano.capolongo@polito.it}


\begin{abstract}
We introduce an operational approach to the use of pulsating sources, located at spatial infinity, for
defining a relativistic positioning and navigation system, based on the use
of four-dimensional bases of null four-vectors, in flat spacetime. As a
prototypical case, we show how pulsars can be used to define such a
positioning system. The reception of the pulses for a set of different
sources whose positions in the sky and periods are assumed to be known
allows the determination of the user's coordinates and spacetime trajectory,
in the reference frame where the sources are at rest.  We describe our approach in flat Minkowski spacetime, and discuss the
validity of this and other approximations we have considered.
\end{abstract}

\begin{keyword}
 relativistic positioning system  \sep astrometry and reference systems \sep null frames.

\end{keyword}

\end{frontmatter}

\parindent=0.5 cm



\section{Introduction}\label{sec:intro}

The current positioning systems, such as GPS and GLONASS \citep{ashby,sanchez}%
, are essentially conceived as \textit{Ptolemaic}, since they are based on a
reference frame centered in the Earth, and \textit{Newtonian}, since
positioning is defined in a classical (Euclidean) space and absolute time,
over which relativistic (post-Newtonian) corrections are introduced \citep
{ashby}. Furthermore, these systems are effective for positioning on the
Earth, but they can hardly be used for navigating in the space outside the
Earth, as in the Solar System and beyond. In contrast, some authors \citep
{coll3,coll4,coll5,rovelli,hehl,ruggiero07,bini08}
recently proposed to use the worldlines of electromagnetic signals, emitted
by objects in geodesic motion, in order to build a relativistic positioning
system, based on the use of the so called  \textit{emission coordinates}. These relativistic positioning systems are also
autonomous or autolocated, since any user can determine its own position
(and spacetime trajectory) by solely elaborating the received signals.

The simplest way of understanding what emission coordinates are, is to
consider four emitting clocks, in motion in spacetime, broadcasting their
proper times: the intersection of the past lightcone of an event with the
worldlines of the emitting clocks corresponds to the proper times of
emission along the worldlines of the emitters; these proper times are the
emission coordinates of the given event. For example, one may think of a set
of satellites orbiting around the Earth and equipped with onboard clocks,
however, such a system would hardly be effective for the navigation in the
Solar System; for that purpose, a set of \textit{pulsars} could rather be
used \citep{collpulsar0,collpulsar}. In fact, known pulsars emit their signals at a highly regular rate
(this is the case, in particular, of the \textit{millisecond pulsars}, see
e.g. \citet{kramer}), which makes them natural beacons for building a
relativistic positioning system. What can be measured with
great accuracy is the arrival time of the N-th pulse, so that counting these
pulses can in principle allow to define something similar to the emission
coordinates, even though in this case the actual proper emission time is unknown and inaccessible.

Actually, the idea of using pulsars as stellar beacons has been
considered since the early years of discovery of pulsars \citep{oldiepuls}, and
other proposals are actually under study \citep{sala,xrays1,xrays2}, which are based
on the accurate measurement of the Times of Arrival (TOA) of the pulses or the phase differences between pulses: in general, in these approaches positioning is not autonomous, but can be referred to
 a reference frame centered at the Solar System barycenter (SSB), because pulses arrival times or their phases have to be related to their expected values in the SSB.

However, thanks to the use of emission coordinates  \citep{collpulsar},  autonomous positioning  by means of pulsars can be defined: in this context, we introduce here an operational approach to the use of periodic sources (assumed to be at rest at spatial infinity)
such as pulsars (anticipated in \citet{corea}), for defining an autonomous  relativistic
positioning and navigation system, based on the use of a four-dimensional
basis of null four-vectors. We assume that a user is equipped with a
receiver that can count pulses from a set of sources whose periods and
positions in the sky are known; then, reckoning the periodic electromagnetic
signals coming from (at least) four sources and measuring the proper time
intervals between successive arrivals of the signals allow to localize the
user, within an accuracy controlled by the precision of the clock he is
equipped with. This system can allow autopositioning with respect to an
arbitrary event in spacetime and three directions in space, so that it could
be used for space navigation and positioning in the Solar System and beyond.
In practice the initial event of the self-positioning process is used as the origin of the reference, and the axes are oriented according to the positions of the distant sources; all subsequent positions will be given in that frame. If one wants to further position the whole section of worldline of the receiver in some other external reference frame\footnote{For instance, in the Solar System events can be described by making use of the Barycentric Celestial Reference System (BCRS).}  the location of the initial event in the external frame has to be known by other means.
We describe our approach in flat Minkowski spacetime, and discuss the
validity of this and other approximations we have considered, for actual physical
situations. The paper is organized as follows: after resuming the basic principles of null frames in Section \ref{sec:grid}, in Section \ref{sec:locgrid} we show how localization takes place in our approach; then, in Section \ref{sec:bias} we discuss the sources of error, \ref{sec:techni} is devoted to the discussion of some practical problems one has to face, while Conclusions are drawn in Section \ref{sec:conc}.


\section{The Basic Null Frame and Definition of the Grid}

\label{sec:grid} 

Let us consider a number of sources of periodic electromagnetic signals, at
rest at spatial infinity, in a four-dimensional Minkowski spacetime. For
our purposes, at least four sources are needed. Each of these sources is
characterized by the frequency of its periodic signals and by their
directions in space; since the sources are supposed to be far away (i.e. at spatial infinity), their
signals can be seen as corresponding to plane waves. In the inertial frame
where the sources are at rest, once Cartesian coordinates are chosen, we
associate to each source a null four-vector\footnote{%
Arrowed bold face letters like $\vec{\mathbf{x}}$ refer to spatial vectors,
while boldface letters like $\bm f$ refer to four-vectors; Greek indices
refer to spacetime components, while Latin letters label the sources.} $\bm f
$ whose Cartesian contravariant components are given by
\begin{equation}
f^{\mu } \doteq \frac{1}{c T}(1,\vec{\mathbf{n}}),  \label{eq:deff}
\end{equation}
$T$ being the (proper) signal period, and $\vec{%
\mathbf{n}}$ the unit vector describing the direction of propagation in
the given frame. If in the same reference frame we consider the position
four-vector
\begin{equation}
\bm{r}\doteq (ct,\vec{\mathbf{x}}),  \label{eq:defr}
\end{equation}%
with respect to an arbitrary and yet unspecified origin, then we can define
the scalar function $X$ at the spacetime event identified by the position
four-vector $\bm r$
\begin{equation}
X(\bm r)\doteq \bm f\cdot \bm r,  \label{eq:defX}
\end{equation}%
where dot stands for Minkowski scalar product. The scalar $X$ might be
thought of as the phase difference of the wave described by $\bm f$ with
respect to its value at the origin of the coordinates. Four linearly independent
four-vectors constitute a basis, or a frame: we may think of choosing four
null four-vectors to serve as a basis (see e.g. \citet{hehl}), so that the
four wave four-vectors $\{\bm f_{(a)},\bm f_{(b)},\bm f_{(c)},\bm f_{(d)}\}$
in the form (\ref{eq:deff}) constitute our \textit{null frame}, or \textit{null
tetrad}. Then, according to the general approach to coordinate systems and frames developed by \citet{coll2009} in connection with positioning systems, the four phase differences
\begin{equation}
X_{(N)}\doteq \bm f_{(N)}\cdot \bm r,\quad N=a,b,c,d\   \label{eq:defXN}
\end{equation}
defined at any event $\bm r$ whose coordinates are defined by (\ref{eq:defr}), with $a,b,c,d$ labeling the sources, are \textit{null coordinates}: in other words, they are spacetime functions with null spacetime gradient and, hence, they define a \textit{null coordinate system}. Furthermore, the $\big\{ X_{(N)} \big \}$ are \textit{emission coordinates}, since they are physically related to the reception of electromagnetic signals emitted by the sources.

In  detail, according to the general tetrad formalism (see
e.g. \citet{chandra}), the null tetrad allows the definition of the symmetric
matrix
\begin{equation}
\eta _{(M)(N)}=\bm f_{(M)} \cdot \bm f_{(N)},  \label{eq:defgab}
\end{equation}%
which, in this case, has constant components, and whose inverse is
determined by the relation
\begin{equation}
\eta _{(M)(P)}\eta ^{(P)(N)}=\delta _{(M)}^{(N)}.  \label{eq:defgabinv}
\end{equation}%
Tetrad indices $N=a,b,c,d$ are lowered and raised by means of the matrices $%
\eta _{(M)(P)}$ and $\eta ^{(M)(P)}$. We can write the position four-vector $\bm r$ in the form
\begin{equation}
\bm r=X^{(N)}\bm f_{(N)}=X_{(N)}\bm f^{(N)},  \label{eq:defrvec}
\end{equation}%
and, as a consequence we see that the phase differences $X_{(N)}$ are the
components of the position four-vector with respect to the vectors
\begin{equation}
\bm f^{(N)}=\eta ^{(N)(M)}\bm f_{(M)},  \label{eq:fcontraN}
\end{equation}%
or, differently speaking, the functions
\begin{equation}
X^{(N)}=\eta ^{(N)(M)} X_{(M)} \label{eq:defXNup}
\end{equation}
are the components of the position four-vector with respect to the null tetrad vectors $\bm f_{(N)}$. It is useful to remark that while the frame $\{\bm f_{(a)},\bm f_{(b)},\bm f_{(c)},\bm %
f_{(d)}\}$ is constituted by null vectors, the frame $\{\bm f^{(a)},\bm f^{(b)},\bm f^{(c)},\bm %
f^{(d)}\}$ is constituted by space-like vectors.

In summary, if we consider the hyperplanes
conjugated to the null frame $\{\bm f_{(a)},\bm f_{(b)},\bm f_{(c)},\bm %
f_{(d)}\}$ vectors, we are able to define a spacetime grid (see \citet{corea}), in
which each event is identified by the relative phase of the electromagnetic
signals with respect to an arbitrary origin and, in this frame, the coordinates of each event are given by the functions $\big\{ X^{(N)} \big \}$; equivalently, the phases $\big\{ X_{(N)} \big \}$ are the coordinates with respect to the space-like frame $\{\bm f^{(a)},\bm f^{(b)},\bm f^{(c)},\bm f^{(d)}\}$.


\section{Localization within the Grid}

\label{sec:locgrid}

After having shown how to build a grid, we want to focus on how localization
can be achieved within the grid. In particular, we suppose to deal with
periodic signals in the forme of electromagnetic pulses, such as those coming from pulsars, and that these signals can be thought of as locally plane waves. Furthermore, we suppose that the user is equipped with a receiver able to recognize and count the pulses coming from the various sources, and a clock, that can be used to measure the proper
time span between the arrivals from each source.

Let us start with a toy model, where the emission from the sources is
continuous and the phases of any signal can be determined with an arbitrary
precision, at any event. We choose a starting event, from which the phases
of each signal are measured, which is the origin of our coordinates (in other
words, the event with $\bm r=\bm0$, according to what we have described
above), and three directions in space, defining the Cartesian axes of the
inertial frame of the sources. We point out that even though the starting
event is arbitrary, in order to correctly define the null frame, the directions
of the sources in the sky have to be known: in other words, we have to know
the unit vectors $\vec{\mathbf{n}}$ for each source (and their proper
frequency $\nu $ too), which also enable us to calculate the matrices $\eta
_{(N)(M)},\eta ^{(N)(M)}$ of the given frame.

To a subsequent event $\bm  r$, we associate the measured phases
\begin{equation}
X_{(N)} = \bm f_{(N)} \cdot \bm r, \quad  \label{eq:defXzero2}
\end{equation}
and, according to eq. (\ref{eq:defrvec}), it is then possible to obtain the
coordinates of the event $\bm r$, in terms of the measured phases:
\begin{equation}
\bm r = X_{(a)}\bm f^{(a)}+X_{(b)}\bm f^{(b)}+X_{(c)}\bm f^{(c)}+X_{(d)}\bm %
f^{(d)}.  \label{eq:defXzero3}
\end{equation}
and to reconstruct the user's worldline.

Coming to a more realistic situation, such as the one in which the emitters
are pulsars, we should consider that the received signals consist in a
series of pulses and are not continuous. In this case, we may proceed as
follows. First, we call \textquotedblleft reception\textquotedblright\ the
event corresponding to the arrival of a pulse from one of the sources. As a
consequence, the position in space-time of an arbitrary reception event can be written in the form
\begin{equation}
\bm r=X_{(a)}\bm f^{(a)}+X_{(b)}\bm f^{(b)}+X_{(c)}\bm f^{(c)}+X_{(d)}\bm %
f^{(d)},  \label{eq:defXzero3bis}
\end{equation}%
with
\begin{eqnarray}
X_{(a)} &=&n_{(a)}+p,  \label{eq:defp} \\
X_{(b)} &=&n_{(b)}+q,  \label{eq:defq} \\
X_{(c)} &=&n_{(c)}+s,  \label{eq:defs} \\
X_{(d)} &=&n_{(d)}+w.  \label{eq:defw}
\end{eqnarray}%
 The $X_{(N)}$'s in the case of a continuous signal would be phases. Here they are given by an integer $n_{(N)}$%
, numbering the order of the successive pulses from a given source, and a fractional value: e.g.
$p$ means a fractional value of the cycle in $X^{(a)}$, and the equivalent
holds for $q,s,w$, where $0<p,q,s,w$; in eqs. (\ref{eq:defp})-(\ref{eq:defw}%
), only one of the $p,q,s,w$ will in general be zero: when for instance a pulse from source \textquotedblleft a\textquotedblright arrives $p$ will be zero, while $q$, $s$, $w$ will not; when a pulse from source \textquotedblleft b\textquotedblright comes $q$ will be zero and the other three fractions will not; and so on. Once we choose an
arbitrary origin, we may count the pulses in order to measure the $n_{(N)}$,
but we have no direct means to measure the fractional values $p,q,s,w$.
However a procedure to determine these values can be obtained, based on
geometric considerations: we suppose that the acceleration of the user is
small during a limited series of reception events, so that we may identify
the user's worldline with a straight line; furthermore, we also suppose that
by means of his own clock the user can measure the proper time interval $%
\tau _{ij}$ between the i-th and j-th arrivals. With these assumptions we
can proceed as follows to determine the fractional values $p,q,s,w$. Let us
consider two sequences\footnote{%
They may be subsequent or not, provided the total time span does not spoil
the hypothesis of linearity of the worldline.} of arrival times from the
sources; we have eight events, each of them in the form
\begin{equation}
\bm r_{j}=X_{(a)j}\bm f^{(a)}+X_{(b)j}\bm f^{(b)}+X_{(c)j}\bm %
f^{(c)}+X_{(d)j}\bm f^{(d)},    j=1,..,8,  \label{eq:arrivalj}
\end{equation}%
where $X_{(N)j}$ are expressions like (\ref{eq:defp}--\ref{eq:defw}). The
events are arranged in such a way that $\bm r_{1}$ is the generic arrival of the
signal from pulsar \textquotedblleft a\textquotedblright , $\bm r_{2}$
is the arrival of the first signal of pulsar \textquotedblleft
b\textquotedblright\ after $\bm r_{1}$, $\bm r_{3}$ is the  arrival of
the first signal of pulsar \textquotedblleft c\textquotedblright\ after $\bm %
r_{1}$, and $\bm r_{4}$ is the  arrival of the first signal of pulsar
\textquotedblleft d\textquotedblright\ after $\bm r_{1}$ (the pulsars are
ordered from largest (\textquotedblleft a\textquotedblright ) to shortest
(\textquotedblleft d\textquotedblright ) period); $\bm r_{5}$ is the arrival
of the next signal from pulsar \textquotedblleft a\textquotedblright , and
so on. The flatness hypothesis allows to write the displacement four-vector
between two reception events in the form
\begin{equation}
\bm r_{ij}\doteq \bm r_{i}-\bm r_{j}=\left( X_{(N)i}-X_{(N)j}\right) \bm %
f^{(N)}\doteq \Delta X_{(N)ij}\bm f^{(N)}.  \label{eq:deltarij}
\end{equation}%
Indeed, the assumption that the worldline of the receiver is straight during
a limited number of pitches of the signals can be used also to provide
further information. In fact, let us consider three successive reception
events i,j,k; we have
\begin{equation}
\bm r_{ji}=\Delta X_{(N)ji}\bm f^{(N)},\quad \quad \bm r_{kj}=\Delta
X_{(N)kj}\bm f^{(N)}.  \label{eq:deltarjirkj}
\end{equation}%
The straight-line hypothesis allows us to write
\begin{equation}
\frac{\tau _{ji}}{\tau _{kj}}=\frac{\Delta X_{(a)ji}}{\Delta X_{(a)kj}}=%
\frac{\Delta X_{(b)ji}}{\Delta X_{(b)kj}}=\frac{\Delta X_{(c)ji}}{\Delta
X_{(c)kj}}=\frac{\Delta X_{(d)ji}}{\Delta X_{(d)kj}},  \label{eq:rapporti1}
\end{equation}%
where $\tau _{ji}$, $\tau _{kj}$ are the proper times elapsed between the
i-th and j-th, and j-th and k-th reception events, respectively. These
relations enable us to obtain the values we are interested in: in fact, we
may arrange the coefficients of eqs. (\ref{eq:arrivalj}) in an $8\times4$ matrix (8 events, 4 sources):
\beq
X_{(N)i}=\left(\begin{array}{c|c|c|c} n_{1}^{(a)} & n^{(b)}_{1}+q_{1} & n^{(c)}_{1}+s_{1} & n^{(d)}_{1}+w_{1} \\\hline n^{(a)}_{2}+p_{2} & n^{(b)}_{2} & n^{(c)}_{2}+s_{2} & n^{(d)}_{2}+w_{2} \\\hline n^{(a)}_{3}+p_{3} & n^{(b)}_{3}+q_{3} & n^{(c)}_{3} & n^{(d)}_{3}+w_{3} \\\hline n^{(a)}_{4}+p_{4} & n^{(b)}_{4}+q_{4} & n^{(c)}_{4}+s_{4} & n^{(d)}_{4} \\\hline n^{(a)}_{5} & n^{(b)}_{5}+q_{5} & n^{(c)}_{5}+s_{5} & n^{(d)}_{5}+w_{5} \\\hline n^{(a)}_{6}+p_{6} & n^{(b)}_{6} & n^{(c)}_{6}+s_{6} & n^{(d)}_{6}+w_{6} \\\hline n^{(a)}_{7}+p_{7} & n^{(b)}_{7}+q_{7} & n^{(c)}_{7} & n^{(d)}_{7}+w_{7} \\\hline n^{(a)}_{8}+p_{8} & n^{(b)}_{8}+q_{8} & n^{(c)}_{8}+s_{8} & n^{(d)}_{8}\end{array}\right) \label{eq:tableXNi}
\eeq
As it can be seen, the $p,q,s,w$ are zero along the main diagonals of the upper and lower square half matrices forming the whole matrix. This happens in correspondence of the arrivals of the pulses from the various sources: on  the arrival of a pulse from \textquotedblleft a\textquotedblright\ the corresponding $p$ is zero, from \textquotedblleft b\textquotedblright\ $q$ is zero, and so on. Then, on using relations like (\ref{eq:rapporti1}) we obtain the fractional values in terms of observed quantities, i.e. proper time intervals measured by the observer. For instance, we have
\begin{eqnarray}
&&p_{1}=0, \quad q_{1}=n^{(b)}_{2}-n^{(b)}_{1}-\left(n^{(b)}_{6}-n^{(b)}_{2} \right) \frac{\tau_{21}}{\tau_{62}},\label{eq:ext111} \\
&&s_{1}=n^{(c)}_{3}-n^{(c)}_{1}-\left(n^{(c)}_{7}-n^{(c)}_{3} \right) \frac{\tau_{31}}{\tau_{73}},\label{eq:ext112}\\
&& w_{1}=n^{(d)}_{4}-n^{(d)}_{1}-\left(n^{(d)}_{8}-n^{(d)}_{4} \right) \frac{\tau_{41}}{\tau_{84}}, \label{eq:ext11} \\
&&p_{2}=n^{(a)}_{1}-n^{(a)}_{2}+\left(n^{(a)}_{5}-n^{(a)}_{1} \right) \frac{\tau_{21}}{\tau_{51}}, \quad q_{2}=0, \label{eq:ext121} \\
&& s_{2}=n^{(c)}_{3}-n^{(c)}_{2}+\left(n^{(c)}_{7}-n^{(c)}_{3} \right) \frac{\tau_{23}}{\tau_{73}}, \label{eq:ext122} \\
&& w_{2}=n^{(d)}_{4}-n^{(d)}_{2}+\left(n^{(d)}_{8}-n^{(d)}_{4} \right) \frac{\tau_{24}}{\tau_{84}}, \label{eq:ext12} \\
&& p_{3}=n^{(a)}_{1}-n^{(a)}_{3}+\left(n^{(a)}_{5}-n^{(a)}_{1} \right) \frac{\tau_{31}}{\tau_{51}}, \label{eq:ext131} \\
&& q_{3}=n^{(b)}_{2}-n^{(b)}_{3}-\left(n^{(b)}_{6}-n^{(b)}_{2} \right) \frac{\tau_{23}}{\tau_{62}}, \label{eq:ext132} \\
&& s_{3}=0, \quad w_{3}=n^{(d)}_{4}-n^{(d)}_{3}+\left(n^{(d)}_{8}-n^{(d)}_{4} \right) \frac{\tau_{34}}{\tau_{84}}, \label{eq:ext13} \\
&& p_{4}=n^{(a)}_{1}-n^{(a)}_{4}+\left(n^{(a)}_{5}-n^{(a)}_{1} \right) \frac{\tau_{41}}{\tau_{51}}, \label{eq:ext141} \\
&& q_{4}=n^{(b)}_{2}-n^{(b)}_{4}-\left(n^{(b)}_{6}-n^{(b)}_{2} \right) \frac{\tau_{24}}{\tau_{62}}, \label{eq:ext142} \\
&&s_{4}=n^{(c)}_{3}-n^{(c)}_{4}-\left(n^{(c)}_{7}-n^{(c)}_{3} \right) \frac{\tau_{34}}{\tau_{73}},\quad w_{4}=0, \label{eq:ext14}
\end{eqnarray}
and so on. Moving the pair of sequences and repeating the elaboration step by step, we are able to reconstruct  the whole worldline of the receiver, in terms of
measured quantities, i.e. proper times.


\section{Biases and uncertainties} \label{sec:bias}

In this Section we will discuss the sources of error and their importance in the positioning  process that we have outlined. Before going into details of our analysis, we would like to preliminarily discuss the nature of errors and inaccuracies that affect our positioning process, in order to clarify how they can be dealt with to improve accuracy.

Roughly speaking we can distinguish between \textit{systematic errors} and \textit{uncertainties} (or \textit{fluctuations}). Systematic errors originate from mismodelling of the physical processes we are dealing with or from poor knowledge of the system parameters; they globally affect the positioning process and are either constant or time dependent. For instance, errors in the angular positions of the sources in the sky as well as in the periods of the pulses are systematic errors;  furthermore, if we deal with pulsars, their angular positions may change because of their proper motion and their periods because of energy loss.  As for the uncertainties, they are related to the stochastic variations of the quantities involved in the positioning process: for instance in the procedure of measuring the arrival times of the pulses, fluctuations are due to the detection device and to the user's clock as well as to the emission mechanism at the surface of the star; in principle also the turbulence of the interstellar plasma could have a role. Systematic errors produce global consequences with an unknown distortion of the re-built spacetime trajectory of the user; the uncertainties transform the worldline in an uncertainty stripe across space-time. Systematics can be reduced by improving our model and the knowledge of its parameters; a statistical analysis and the best technologies can help to reduce the impact of random disturbances.

\subsection{Model Limitations} \label{ssec:modelim}

Let us begin by discussing the validity of our model, which is based on the propagation of electromagnetic signals originating from pulsating sources  at rest in a given reference frame, in Minkowski spacetime.  The question we would like to address is: how realistic is this model? Can it be used for positioning in actual physical situations? We start with some general observations.

We explicitly work in flat spacetime, thus eliminating the effects of the gravitational field. It is however obvious that, for positioning in the Solar System, the gravitational field of the Sun (and of the other major bodies) influences the propagation of electromagnetic signals.\footnote{En passant, we notice that since in our procedure only ratios of proper times measured in practice at the same spatial position are involved, the effect of the gravitational field on the user's clock can be safely neglected.}
The dimensionless magnitude of the static gravitational field of the Sun is of the order of $\delta_{\odot} \simeq \frac{GM_{\odot}}{c^{2}d} \simeq 10^{-8} \left(\frac{1\ \mathrm{A.U.}}{d} \right)$\footnote{The distance $d$ is here expressed in astronomic units (A.U.) so that a distance equal to $1$ corresponds to the position of the Earth.}, and reaches its maximum value near the Sun, where $\delta_{\odot} \simeq 10^{-6}$. This field produces effects on the times of arrival of the pulses which are relevant for our purposes only if they change in a time comparable with the integration times used for our algorithm; only the radial component (with respect to the Sun) is important. The effects due to the motion of a user who travels with speed $\vb{v}$ in the radial direction $\hb r$ (and thus experiences a time varying gravitational field) in a time span $\delta t$ are expressed in terms of apparent fractional change of the period of the sources as $\delta_{\odot,v} \simeq  \frac{GM_{\odot}}{c^{2}d^{2}} {\vb{v}\cdot\hb{r}}\  \delta t \simeq 6 \times 10^{-8}  \left(\frac{1\ \mathrm{A.U.}}{d^{2}} \right)  \left(\frac{v}{\mathrm{30 \ km/s}} \right) \mathrm{y}^{-1} \delta {t} $\footnote{The velocity is in units of the average speed of the Earth along its orbit (30 km/s) and the rate of change is per year ($\mathrm{y}^{-1}$).}. This fixes an upper limit to the acceptable $\delta t$ in order the effect to be compatible with the required tolerance. One more problem is that when the line of sight to a source grazes the Sun the time of flight of the signals depends on the geometric curvature of the rays and on the Shapiro time delay, which is indeed huge depending on the apparent impact parameter \citep{straumann04}. However these disturbances can be dealt with either choosing sources which are located in the sky so that their lines of sight are far away from the gravitating bodies; or having recourse, which is appropriate for many other reasons also, to a redundant number of sources, so that only the best are used each time.

Another systematic error comes from considering the sources as being at rest, at spatial infinity, in a given reference frame,
which is a very idealized situation. In fact, actual sources, such as galactic pulsars, have both a proper motion and a finite distance from the observer. Taking into account estimates of the proper motion of real pulsars \citep{proper} it is possible to see that  the rate of change of the angular position\footnote{Actually, proper motion affects any observed change in periodicity, see e.g. \citet{straumann04}, \citet{damour91}, so that an estimate of the relative time variation of the period is of the order $10^{-10} \left(\frac{\mathrm{100 \ pc}}{d}  \right)\mathrm{y}^{-1} \delta t$. } is of the order $10^{-6} \left(\frac{ \mathrm{100 \ pc}}{d} \right)\mathrm{rad }$ per year. In practice these figures tell us that we are allowed to keep the position nominally fixed for months before correcting the value of the direction cosines, which is possible because the behaviour of the sources is known. As for the radial proper motion of the pulsar, which is commonly far worse known than its transverse motion, it is indeed contained in the times of arrival of the pulses, but can reasonably be considered as fixed during very long times, practically not affecting the positioning process.
Finally it is known that the periods of the pulses are generally increasing with time, due to rotation energy loss of the star. However the decay rate is in general known, so that we may deal with this kind of change in the same way as we do for the position in the sky: keep the period fixed for a time compatible with the required accuracy (which would in any case be rather long), then introduce a corrected new value (which is known once the pulsar is given). As for glitches (sudden and random jumps in the frequency) they can be cured thanks to the redundancy of the number of sources above the minimum of four, corresponding to the dimensions of spacetime.

In the case of pulsars there are many other effects that could affect the arrival times of the signal; a thorough analysis can be found in \citet{kramer}. Besides what we have already mentioned, we should, for instance, take into account the emission mechanism at the surface of the star and the propagation across the interstellar plasma: these effects are responsible for the fact that each single incoming pulse is in general different from any other. The method to deal with this is the same adopted at radiotelescopes: the acquisition of the data runs for a high number of pulses, then the signal is analyzed by folding, in order to extract the fiducial sequence with the desired accuracy. In the case of the positioning far shorter integration times are needed, because the pulsar is already known and the aim is simply to recognize it, rather than investigate on it. For completeness we should finally mention the problems related to the acquisition and elaboration chain in the receivers, however to discuss the technological aspects of the antenna and acquisition apparatus is beyond the scope of the present work.

Finally, we would like to focus on the approximation that is implicit in our method for the conversion of a sequence of times of arrival of discrete pulses into the coordinates of the receiver.
We exposed the method in Section \ref{sec:locgrid}, where we limited the extension of the time series of the signals we used at each step of the process to an interval allowing to locally approximate the user's worldline with a straight line:
how far is this hypothesis tenable? Given the user's clock accuracy $\delta \tau $ \footnote{Into this $\delta \tau$ we should actually include also the drifts due to the proper motion and the period decay of the pulsar, which we keep constant during one step of the process; the latter are however expectedly far smaller effects than those due to the acceleration of the receiver.},
we can define the maximum proper time interval $\Delta \tau _{\mathrm{max}}$ that can be considered in order to be self consistent with the straight-line hypothesis. Developing the worldline of the receiver in powers of its
proper time up to the second order we see that, if $a$ is the order of
magnitude of the user's acceleration, and $v$ his velocity, the following
condition should be satisfied:
\begin{equation}
\Delta \tau _{\mathrm{max}}=\sqrt{2\frac{v}{a}\delta \tau }.
\label{eq:approx2}
\end{equation}%
For instance, if the user is moving in flat spacetime with $\delta \tau \simeq 10^{-10}$ s, $v=5\times 10^{5}$
m/s and an acceleration $a=1$ m/s$^{2}$, we have $\Delta \tau _{\mathrm{max}%
}=10^{-2}$ s, which corresponds to several periods of millisecond pulsars: enough both for the averaging away of the fluctuations of the single pulses, and for the piecewise reconstruction of the worldline.
Actually, the deviation from the linearity of the user's
worldline can also be due to the curvature of spacetime, i.e. to the
presence of the gravitational field. We can give a similar estimate of the
corresponding maximum proper time interval $\Delta \tau _{\mathrm{max}}$ by
setting $a=|\nabla \Phi| $ in (\ref{eq:approx2}) where $\Phi $ is the
gravitational potential. For instance, for $a=10^{-3}$ m/s$^{2}$, which is
the order of magnitude of the gravitational field of the Sun at 1 A.U., we get for $%
v=10^{3}$ m/s, $\Delta \tau _{\mathrm{max}}\simeq 10^{-2}$ s.
So, we see that there are actual physical situations where the hypothesis of linearity
holds, and the procedure that we have described is meaningful.\\

\subsection{Errors in the procedure of position determination} \label{ssec:souerr}

After having discussed the systematic errors related to the physical model underlying our procedure of position determination, we would like to turn to the analysis of the errors in the procedure itself which, as we are going to see,  are connected both with our knowledge of the system parameters and with the measurement process.

Actually,  the coordinates of an event $\bm r$ are determined by solving  eq. (\ref{eq:defXzero2}). We notice that eq. (\ref{eq:defXzero2}) can be written in the form
\beq
A \overline{x}=\overline{y} \label{eq:syslin1}
\eeq
where\footnote{We use $\overline{x}$ to refer to generic vectors belonging to $\mathbb{R}^{4}$, not to be confused with the four-vectors.}
\beq
A=\left(\begin{array}{cccc}f_{(a)}^{0} & -f_{(a)}^{1} & -f_{(a)}^{2} & -f_{(a)}^{3} \\f_{(b)}^{0} & -f_{(b)}^{1} & -f_{(b)}^{2} & -f_{(b)}^{3} \\f_{(c)}^{0} & -f_{(c)}^{1} & -f_{(c)}^{2} & -f_{(c)}^{3} \\f_{(d)}^{0} & -f_{(d)}^{1} & -f_{(d)}^{2} & -f_{(d)}^{3}\end{array}\right), \quad  \overline{x}=\left(\begin{array}{c}x^{0} \\x^{1} \\x^{2} \\x^{3}\end{array}\right), \quad \overline{y}=\left(\begin{array}{c}X_{(a)} \\X_{(b)} \\X_{(c)} \\X_{(d)}\end{array}\right)  \label{eq:defA1}
\eeq
with $x^{0}=ct$, $x^{1}=x$, $x^{2}=y$, $x^{3}=z$.
Eq. (\ref{eq:syslin1}) is a linear system where the unknown vector $\overline x$ is obtained in terms of the matrix $A$ (which has to be non singular) and the vector $\overline y$:

\beq
\overline{x}=A^{-1} \overline{y}. \label{eq:syslininv}
\eeq

 The entries of matrix $A$ are related to the signal periods $T_{(N)}=(c f_{(N)}^{0})^{-1}$ and the direction cosines $n^{i}_{(N)}=cT_{(N)}f^{i}_{(N)}$ defining the angular positions of the sources; the fact that the $\vb{n}_{(N)}$'s, $N=a,b,c,d$ are different ensures that the matrix is not singular. As for the vector $\overline y$, its entries are the measured phases  determined (at each event) by the procedure described in Section \ref{sec:locgrid}.

The entries of the matrix $A$ are affected by systematic errors, while phase measurements have random errors. It is then possible to evaluate the maximum relative error in the solutions $\overline{x}$ of the linear system (\ref{eq:syslininv}) by estimating the maximum errors on the entries of $A$, while a covariance analysis allows to quantify how random errors in phase measurements translate into errors in  the determination of the coordinates of the spacetime event $\bm r$, on the basis of the geometric properties of matrix $A$.

In fact,  after defining a suitable norm\footnote{For instance: $\|A \|=\sqrt{\sum_{i,j}|a_{ij}|^{2}}, \quad \|\overline x \|=\sqrt{\sum_{i}|x_{i}|^{2}}$. }, the following relation holds (see e.g. \citet{numerical})
\beq
\delta \overline x \leq k(A)^{2} \delta A  \label{eq:errmax2}
\eeq
where  $\delta \overline x = \frac{\| \Delta \overline x \|}{\| \overline x \|},  \delta A = \frac{\| \Delta A \|}{\| A \|}$ and $k(A)= \|A \| \|A^{-1} \|$ is the condition number of the system (\ref{eq:syslin1}).
If we suppose that the relative errors in periods $\Delta T /T$, and direction cosines $\Delta n^{i}$ are roughly the same for all sources in all directions, then eq. (\ref{eq:errmax2}) allows to conservatively estimate the relative error in the form
\beq
\delta \overline x  \leq k(A)^{2} \left[ \sqrt{ \left(\frac{\Delta T}{T} \right)^{2}+\frac 3 2 \left(\Delta n\right)^{2}}
  \right]  \label{eq:x1}
\eeq
We remember that it is always $k(A) \geq 1$; in our case, we may write
\beq
k(A) \propto  \frac{1}{|\mathrm{det}(A)|^{2}} \sum_{N=a,b,c,d}^{4}\frac{1}{T^{2}_{(N)}c^{2}}\label{eq:kaprop1}
\eeq
As a consequence, we see that the relative error is minimized when: (i) the determinant is maximized and (ii) periods are minimized; since the spatial components of $A$ are the direction cosines, the determinant is maximized when the volume spanned by these directions is maximized. To fix ideas, for $\Delta T/T \simeq 10^{-4}, \ \Delta n \simeq 10^{-8},$ it is $\delta x \lesssim k(A)^{2} 10^{-4}$. \\

Further insight on the accuracy achievable by our positioning procedure can be obtained by discussing the geometric properties of the system (\ref{eq:syslininv}), which ultimately depends on the actual angular positions of the sources in the sky: we refer to the Geometric Dilution Of Precision (GDOP), which is based on the covariance matrix of the errors in position determination, and provides a measure of how fit the set of sources is: indeed,  it is related to the geometric  properties of the matrix $A$ and a great accuracy can be obtained if sources are chosen that are sufficiently scattered in the sky (this is pretty much like what happens with GPS satellites, see e.g. \citet{gps}).
The covariance matrix of the positioning errors, as determined by the phase measurements, can be expressed as
\beq
\mathrm{cov} \ \overline{x}= \left[A^{T} \left(\mathrm{cov} \ \overline{y}\right) A \right]^{-1} \label{eq:cov1}
\eeq
If the phase measurements errors from each source are uncorrelated and Gaussian distributed, with zero mean and  variances $\sigma_{(N)}^{2}$, the position covariance matrix turns out to be
\beq
\mathrm{cov} \ \overline{x}= \left[ A^{T} \mathrm{diag}(\sigma^{2}_{(a)},\sigma^{2}_{(b)},\sigma^{2}_{(c)},\sigma^{2}_{(d)}) A\right] ^{-1} \label{eq:cov11}
\eeq
The GDOP is defined by
\beq
\mathrm{GDOP}=\sqrt{\mathrm{Tr}(\mathrm{cov} \ \overline{x})} \label{eq:GDOP1}
\eeq
To understand the meaning of GDOP, let us consider the case when all phase measurements have the same variance $\sigma^{2}$, then $\mathrm{cov} \ \overline{y}=\mathrm{diag}(\sigma^{2}_{},\sigma^{2}_{},\sigma^{2}_{},\sigma^{2}_{})$; furthermore it is $(A^{T}A)^{-1}=\frac{1}{|\mathrm{det}A|^{2}} \left(\mathrm{cof} A\right)^{T}\left(\mathrm{cof} A\right)$ so that we can write
\beq
\mathrm{GDOP}= \frac{\sigma}{|\mathrm{det}A|} \sqrt{\mathrm{Tr\left(\mathrm{cof} A\right)^{T}\left(\mathrm{cof} A\right)}} \label{eq:GDOP2}
\eeq
Again, we see the role of $|\mathrm{det}(A)|$ in minimizing the errors:  GDOP is minimized  when the solid angle spanned by the directions of the sources is maximized.

The  error in position determination can be minimized by using a number $n$ of sources greater than four, which will result in an overdetermined linear system in the form  (\ref{eq:syslin1}), $B \overline x=\overline y$,   provided that now $B$ is an $n\times4$ matrix and $\overline y$ is a vector of $n$ measured phases. Redundant equations allow to improve the accuracy of the solutions via least squares estimation techniques (see e.g. \citet{LS}), so that Eq. (\ref{eq:cov1}) is generalized to\footnote{Notice that if $n=4$ it reduces to Eq. (\ref{eq:cov1}).}
\beq
\mathrm{cov} \ \overline{x}=(B^{T}B)^{-1}B^{T}\left(\mathrm{cov} \ \overline{y}\right) B( B^{T}B)^{-1}   \label{eq:coverrLSe}
\eeq
If, as before,  we consider the phase measurements errors to be identically Gaussian distributed with the same variance and independent, we obtain equation (\ref{eq:cov1}) back, also in the case $n>4$: in summary, we may say that the components of $(B^{T}B)^{-1}$ determine how phase errors translate into errors of the computed spacetime position.


\section{Technical problems}\label{sec:techni}

The purpose of the present paper is not to design a working navigation system, rather to describe a new method allowing relativistic positioning when a set of pulsed sources of electromagnetic signals and an onboard clock are available. We refer mostly to pulsars because they represent natural sources of regular pulses and they are suggestive of the old navigation at sea using the stars; an advantage of using pulsars is the fact that they can practically be considered as being at infinite distance and occupying fixed positions in the sky\footnote{Actually this is not the case, but we have already commented on how to treat the issue in Section \ref{ssec:modelim}.}, however a major drawback is the weakness of their signals. Being aware of this, we have gone on with our example of pulsars without facing the practical issue of the detection of the signals, but our method can perfectly work with any other artificial source of pulses provided one knows the law of motion of the source in a given reference frame. For the use in the Solar system, one could for instance think to lay down regular pulse emitters on the surface of some celestial bodies: let us say the Earth, the Moon, Mars etc. The behaviour of the most relevant bodies is indeed pretty well known, so that we have at hands the time dependence of the direction cosines of the pulses: this is enough to apply the method and algorithm we have described and the final issue in this case would be the position within the Solar system. In principle the same can be done in the terrestrial environment: here the sources of pulses would be onboard satellites, just as it happens for GPS, but without the need of continuous intervention from the ground: again the key point is a very good knowledge of the motion of the sources in the reference frame one wants to use.

In the previous sections we have many times mentioned the fact that the number of independent sources to be used is at least four. The reason for that number is that the space-time coordinates needed to localize the user are of course four (three space components plus time). There are however many reasons for using a bigger number of emitters. If we have $n>4$ pulsars (or equivalent sources) we may apply our method to all possible quadruples contained in $n$: the position will be determined as an average of the results obtained from all quadruples. This would be one more way to average the effect of random disturbances at the emission of the pulses out. Furthermore if one of the sources fails or disappears for any reason (e.g. because it is eclipsed) the localization process is not interrupted; if one new source comes in, provided its position in the sky and proper recursion time are known, the sequence of its pulses is hooked to the main sequence of the arrival times and may be used for further positioning. What matters is that all arrival times from all visible sources are arranged in one single sequence identified by the onboard clock and that the main sequence is not interrupted. To say better, the maximum tolerable total blackout of the sequence should not last more than what can be reconstructed by extrapolation from the last portion of the world-line of the observer. With reasonable accelerations at the typical speeds of an interplanetary travel this may be as long as a few seconds.

Of course everything works if the clock used by the observer is a good and reliable clock. We are considering atomic clocks fit for being carried on board a spacecraft. The problems with the clocks arise from both random and systematic instabilities. The former contribute to the general uncertainty in the positioning, whilst the latter introduce an apparent shortening of the period of the received pulses. A strategy to minimize the effects of the drift in the frequency of the clock is similar to the one adopted for GPS. It is redundancy: more than one atomic clocks, and of different types, should be carried by the observer (GPS satellites have four clocks on board): the arrival times would be given by an average of the different readings. The redundancy would partly reduce also the effect of the drift of the frequency of the clocks, but this would emerge in long time. The accuracy of portable present day atomic clocks is in the order of $1\times10^{-14}$ (see e.g. \citet{ashby}) which means that the measured repetition time of a millisecond pulsar would be significantly affected by the frequency drift of the clock over times of the order of months, if not years. It would in any case be advisable to periodically check the accuracy of the measurement from some reference station on the Earth, and the best strategy would be to check the position and compare with the one obtained by self positioning, rather than to try to directly verify the clock, which would be possible but it is a very delicate task indeed and presupposes, any way, a good knowledge of the relative motion between the space observer and the ground station.
A periodic check up (after weeks or months intervals) would in any case be cheaper than a continuous guidance from the ground.

Coming back to the initial lines of the present section, we stress again the fact that ours is not the design of a working device,  but rather the description of a principle method to attain self positioning by means of the signals emitted by periodic sources from known positions in the sky. In the same spirit we have here shown that we are considering also specific strategies to tackle some relevant  practical problems one would have to face when coming to implement our proposal in the real world.


\section{Conclusions}

\label{sec:conc}

We have described an operational approach to the use of pulsating signals for positioning purposes; in particular, pulsars signals can be used.
Our procedure is based on the definition of a null frame, by means of the four-vectors associated to the signals in the inertial reference frame where the sources are at rest
(so that the emission directions and the frequencies of the pulsating
signals have to be known) and far away (so that their signals can be dealt
with as plane waves). The procedure is fully relativistic and allows position
determination with respect to an arbitrary event in flat spacetime. Once a null frame has been
defined, it turns out that the phases of the electromagnetic signals  can be
used to label an arbitrary event in spacetime. If the
sources emit continuously and the phases can be determined with arbitrary
precision at any event, it is straightforward to obtain the coordinates of
the user and his worldline. However, actual sources emit signals which consist of a
series of pulses: so, we have developed a simple method that can be used to
determine the user's worldline by assuming that the
worldline is a straight line during a proper time interval corresponding to
the reception of a limited number of pulses, which means that the effects of the
acceleration are negligibly small.

In discussing the source of errors which affect the positioning process, we considered
systematic errors, due to mismodelling or to the poor knowledge of the system parameters, and uncertainties, due to the stochastic changes of the quantities involved in the procedure.
While the former introduce global consequences resulting in an unknown distortion of the reconstruction of the user's worldline and can be reduced by improving our model and the knowledge of its parameters, the latter transform the worldline in an uncertainty stripe across spacetime and can be reduced by means of a statistical analysis and an improvement of the technologies involved. We have discussed  the effects of the gravitational field, of the motion of the sources and of the variability of their emission, the validity  of the hypothesis of linearity, and we have shown that their impact is small enough or can be corrected  for a number of actual physical situations, so that the procedure that we have described can be safely applied. Also, we have evaluated the accuracy that can be achieved, by taking into account the maximum errors in the parameters of the system and by quantifying the impact of random  measurement errors on the determination of the spacetime trajectory, introducing the Geometric Dilution Of Precision.

We did not enter into several technological issues that are important for building an actual positioning system based on our procedure, for instance
the extreme weakness of signals from pulsars and the required design and
sensitivity of the detectors. These and other issues require further
investigations and perhaps technological improvements, however we believe
that this approach could be useful for defining an autonomous and
relativistic positioning system in space, which ultimately transfers the
basic positioning frame from the Earth to spacetime, according to a truly
relativistic viewpoint.


\section*{Acknowledgements}

\label{sec:ack} 

Our research has been supported by Piemonte local government within the
MAESS-2006 project "Development of a standardized modular platform for
low-cost nano- and micro-satellites and applications to low-cost space
missions and to Galileo" and by ASI. We  thank Dr. R. Molinaro, who collaborated during
the early stages of this work.





\clearpage

\end{document}